# Unpinning the skyrmion lattice in MnSi; the effect of substitutional disorder

C. Dhital[1, 2*], L. DeBeer-Schmitt[3], D. P. Young[1] and J. F. DiTusa[1**]

*1Department of Physics and Astronomy, Louisiana State University, Baton Rouge, LA 70803*

*2Department of Physics, Kennesaw State University, Marietta, GA, 30060*

*3Oak Ridge National Laboratory, Oak Ridge, Tennessee 37831, USA*

## Abstract:

By employing magnetization and small angle neutron scattering (SANS) measurements, we have investigated the behavior of the skyrmion lattice (SKL) and the helical order in $MnSi_{0.992}Ga_{0.008}$. Our results indicate that the order of the SKL is sensitive to the orientation of an applied magnetic field with respect to the crystal lattice and to variations in the sequence of small temperature and applied magnetic field changes. The disorder caused by the substitution of the heavier element Ga for Si is sufficient to reduce the pinning of the SKL to the underlying crystalline lattice, reducing the propensity for the SKL to be aligned with the crystal lattice. This tendency is most evident when the applied field is not well oriented with respect to the high symmetry axes of the crystal resulting in disorder in the long range SKL while maintaining sharp short range (radial) order. We have also investigated the effect of substituting heavier elements into MnSi on the reorientation process of the helical domains with field cycling in $MnSi_{0.992}Ga_{0.008}$ and $Mn_{0.985}Ir_{0.015}Si$. A comparison of the reorientation process in these materials with field reduction indicates that the substitution of heavier elements on either Mn or Si sites creates a higher energy barrier for the reorientation of the helical order and for the formation of domains.

## 1. Introduction

The antisymmetric interactions between magnetic moments allowed by the broken space inversion symmetry in non-centrosymmetric chiral materials, such as *B20* compounds, stabilize a topologically non-trivial spin texture known as a magnetic skyrmion lattice (SKL) [1-2]. The periodicity and chirality of the SKL, formed by the superposition of three helices, are determined primarily by the Dzyalloshinskii-Moriya interaction (*D*) in combination with the uniform exchange interaction (*J*) [1-3]. The winding number of these spin textures differentiates them from other trivial magnetic textures such as bubbles thereby providing topological protection [2-4] making the SKL an ideal candidate for low power applications in novel spintronic and information storage devices [5-6]. These unique magnetic textures have been observed in several chiral structured magnets such as MnSi, FeGe, MnGe, $Cu_2OSeO_3$, Co-Mn-Zn, $Fe_{1-x}Co_xSi$, $Mn_{1-x}Co_xSi$, and $Mn_{1-x}Fe_xSi$ [1,2,7-11]. The magnetic phase diagrams of these chiral magnets are qualitatively similar with a helimagnetic zero field ground state, a conical magnetic state at low applied field, a SKL or A-phase for moderate fields and temperatures approaching the critical point, and a high field polarized magnetic phase [1, 2, 8, 9,12,13,14]. However, many of the features, such as the size, chirality, orientation of the helices or the SKL, the magnetic transition temperature, $T_C$, and the magnitude of the ordered moment vary widely and are all material specific [1,2,7,8,9,13,14].

The formation and orientation of the helices and skyrmion lattice are described by the Landau free energy functional of the form $f=f_0+f_{anisotropic}$ [1-3]. The first term $f_0$ contains isotropic terms and includes the contributions from uniform exchange interaction ($J$), the Dzyalloshinskii-Moriya interaction ($D$), and the Zeeman energy of the form $\boldsymbol{H} \cdot \boldsymbol{M}$ [1,2]. This term, $f_0$, where

$$f_0[M] = \int d^3r \left( r_0 M^2 + J(\nabla M)^2 + 2DM \cdot (\nabla \times M) + UM^4 - \mu_0 H \cdot M \right), \qquad (1)$$

is sufficient to describe the formation of a helical magnetic state with a wave vector $Q \sim D/J$. Here, $H$ is the external magnetic field and $r_0$, $J$, $D$, $U$ are phenomenological parameters with ($U$, $J > 0$). $U$ is the mode coupling parameter and $r_0$ represents the distance from the mean field magnetic transition temperature. The last term $\boldsymbol{H} \cdot \boldsymbol{M}$ is the Zeeman energy in the presence of an external magnetic field with $\boldsymbol{M}$ the local magnetization. For MnSi, $D>0$, which gives a left (right) handed spiral for a left (right) handed crystal structure with wave vector $Q=|\boldsymbol{Q}| \sim D/J$ [1-3]. Although, the term $f_0$ is sufficient to describe the formation of helical magnetic order, the orientation of the helix and a SKL with respect to the crystal lattice is governed by terms containing anisotropic interactions $f_{anisotropic}$ [1, 2, 3, 8, 15-18], with

$$f_{anisotropic} = \epsilon_1 \left( \hat{Q}_x^4 + \hat{Q}_y^4 + \hat{Q}_z^4 \right) + \mathcal{O}(\delta^4) + \epsilon_2 \left( \hat{Q}_x^2 \hat{Q}_y^4 + \hat{Q}_y^2 \hat{Q}_z^4 + \hat{Q}_z^2 \hat{Q}_x^4 \right) + \mathcal{O}\left( \delta^6 \right). \qquad (2)$$

The terms containing $\epsilon_1$ and $\epsilon_2$ vary as fourth and sixth order of the spin orbit interaction strength, $\delta$, respectively since $D \propto \delta$. The other fourth order terms $\mathcal{O}(\delta^4)$ i.e. $(\hat{Q}_x^2 \hat{Q}_y^2 + \text{cyclic terms})$ are redundant. There are other sixth order terms $\mathcal{O}(\delta^6)$ i.e $(\hat{Q}_x^2 \hat{Q}_y^2 \hat{Q}_z^2)$ and $\hat{Q}_x^6 + \text{cyclic terms}$), that preserve the $C_4$ symmetry and are less important for a six fold symmetric SKL. The coefficients $\epsilon_1$ and $\epsilon_2$ represent the strength of the magneto-crystalline anisotropy [1,2,3,8,16-18]. The orientation of the helix in zero field is determined by the fourth order term with coefficient $\in_1$, and the orientation of the SKL is dictated by the sixth order term with coefficient $\in_2$. For practical applications, details such as the size, chirality, orientation, $T_C$, and magnitude of the ordered moment are all very important as they determine the necessary conditions for manipulating the SKL. For example, the potential that pins the SKL to a particular crystallographic orientation dictates the value of critical current required for translation [16, 19, 20]. Therefore, it is highly desirable to explore the behavior of these magnetic lattices under different physical and chemical environments in order to probe the mechanism responsible for material specific variations.

Previous small angle neutron scattering (SANS) studies have indicated that the orientation of the helical ordering is along the [111] for MnSi [1,2], along either the [110] or [001] for $Fe_{1-x}Co_xSi$ depending on $x$[8], and along the [001] for $Cu_2OSeO_3$ [9]. Furthermore, the orientation of the helix in FeGe varies from the [111] to the [001] with temperature [10,14, 21,22]. It is not well understood why such variation occurs among these materials with same crystal environment and why it may be dependent on temperature and composition. These differences may originate in details such as the sign of $\in_1$ and $\in_2$, and the variation in the spin-orbit coupling strength. However, the orientation of the SKL closely follows the orientation of an applied magnetic field and displays significant temperature and magnetic field history dependence in the presence of disorder [17, 18,23]. Furthermore, SANS studies of MnSi have yielded the following observations about the orientation and the field and temperature history dependence of the SKL.

1. The SKL pins preferentially along a [110] equivalent provided this direction lies close to the plane perpendicular to the magnetic field. If there are no [110] equivalents close to this plane, then the SKL tends to pin along a [100] [1].
2. When more than one [110] equivalent crystallographic direction is available in the plane perpendicular to the magnetic field, multiple domains of the SKL can form [1,17].
3. The reorientation of the helical lattice upon field cycling or rotation in a magnetic field is elastic. That is, there is a tendency for the helical state to form domains along all equivalent [111] directions following a field cycling into the field polarized state [18].

However, the following questions have not been addressed.

1. What is the orientation of the SKL when both [110] and [001] directions are absent in the plane perpendicular to the magnetic field and how does the SKL respond to a sequence of field and temperature variations in this case?
2. What changes occur to the long range (orientational) order of the SKL and the helical order due to substitutional disorder sufficient to significantly modify $T_C$?
3. What effect does substitutional disorder have on the elasticity of magnetic domains, does it depend on which of the two lattice sites is disordered, and what is the cause for domain formation in helimagnets?

The answers to these questions are important for developing an understanding of the mesoscopic properties of the SKL and for future practical applications in spintronics. In addition, the answers will provide insight into the effect of heavy impurity atoms on the formation, orientation and reorientation process of SKL and helices in the presence of disturbing fields, or temperature variations.

To address these questions, we have performed detailed magnetization and small angle neutron scattering (SANS) measurements of $MnSi_{0.992}Ga_{0.008}$ single crystals. The effect of variations of the temperature, field magnitude, and field orientation, and their history dependencies, on the magnetic structure was explored. The Ga substitution results in an expansion of the crystal lattice sufficient to raise the $T_C$ by 5 K [25]. Our observations indicate that the pinning of the helix along a [111] direction (easy axis) remains robust to the disorder induced by the Ga substitution, whereas the orientation of the SKL can be more easily manipulated. The orientation of the SKL can switch easily from aligning with a [110] to a [100] (hard axis) direction by varying the field or changing the thermal history. We find that in the case where neither a [110] or [100] direction are adjacent to the plane perpendicular to the magnetic field, a ring of scattering appears on the detector instead of the iconic hexagonal scattering pattern of the prototypical skyrmion lattice. This indicates either an orientational disorder of the skyrmions, the presence of a large number of randomly oriented SKL domains, or the existence of labyrinth domains [24] whenever the field is not within 10 degrees of perpendicular to a high symmetry direction of the crystal.

In addition, we have probed the effect of heavy element substitution into MnSi on the elasticity of the helical ordering in $MnSi_{0.992}Ga_{0.008}$ [25] and $Mn_{0.985}Ir_{0.015}Si$ [26] single crystals. Within the time scale of our measurement, the orientation of the helical state recovers (along the equivalent [111] directions) after field cycling with a decreased intensity for $MnSi_{0.992}Ga_{0.008}$. In contrast, we observe no evidence for such recovery in the case of $Mn_{0.985}Ir_{0.015}Si$. This indicates that the elasticity of helical lattice reorientation decreases with substitution of heavier elements on either lattice site, perhaps due to the strength of the disorder and/or the expected increase in spin orbit coupling.

## 2. Experimental Details

### I. Crystal growth and magnetic characterization.

Single crystals of $MnSi_{0.992}Ga_{0.008}$ were grown using Ga flux as described in Ref. [27]. The resulting crystals were pyramidal shaped having a mass of up to 70 mg. Single crystals of $Mn_{0.985}Ir_{0.015}Si$ were grown via a modified Bridgeman technique. The chemical composition of these crystals was determined using Wavelength Dispersive Spectroscopy (WDS). The obtained crystals were single phase crystals having the cubic *B20* crystal structure as determined from single crystal and powder X-ray diffraction. We have also confirmed that the $MnSi_{0.992}Ga_{0.008}$ crystal used in the SANS experiments presented in this paper was single domain via measurements performed on HB3A, a four-circle neutron diffractometer at the High Flux Isotope Reactor (HFIR) at Oak Ridge National Laboratory (ORNL). For $MnSi_{0.992}Ga_{0.008}$, the Ga substitution increases the lattice constant by ~ 0.1% creating negative chemical pressure thereby increasing $T_C$ and modifying the bulk magnetic behavior [25]. The residual resistivity ratio of these crystals was approximately 60 as reported in Ref. [25]. Whereas for Ir substitution, we observe a systematic decrease of $T_C$ along with an increase in $Q$ [26]. ac and dc magnetization measurements were carried out in a Quantum Design 7-T Magnetic Property Measurement System (MPMS) superconducting quantum interference device (SQUID) magnetometer. ac susceptibility ($\chi'$ *and* $\chi''$) was measured by applying an ac signal on top of the dc magnetic field. More details of the magnetic properties of these samples are presented in our previous publications [25,26].

### II. Small angle Neutron scattering (SANS).

SANS measurements were carried out at the GP-SANS beamline at HFIR. All of the crystals were aligned such that the [-110] crystal direction was along the magnetic field which was oriented parallel to incident beam. In addition, the crystalline [111] direction was oriented nearly vertical such that it lies within the detector plane. The mean wavelength of incident neutrons employed was $\lambda = 6$ Å with $\Delta\lambda/\lambda = 0.13$ and a sample to detector distance of 11.3 m. The measurements were performed under a set of field and temperature sequences that are summarized in the figures and the movie in the supplementary information [28]. In addition, for $MnSi_{0.992}Ga_{0.008}$, a set of measurements were performed after a rotation of the crystal about the vertical (the [111] direction) for angles between 0 and $\pi/2$ while keeping the field and beam directions constant. Slight variations of the [111] reflection as a function of rotation are indicative of a misalignment of the sample of a few degrees. Similar measurements were performed on a $Mn_{0.985}Ir_{0.015}Si$ single crystal with the exception of crystal rotation experiments.

### 3. Results

**I. Magnetic Properties.**

The dc and ac magnetization of a representative $MnSi_{0.992}Ga_{0.008}$ crystal are shown in Fig.1. Fig.1a displays a comparison of the magnetization divided by $H$ (*M/H*) of MnSi and $MnSi_{0.992}Ga_{0.008}$ single crystals both measured at 1 kOe. The substitution of a small amount of Ga into MnSi has increased $T_C$ by ~5 K over that of nominally pure MnSi ($T_C$~28.5 K) due to the negative chemical pressure created by the Ga substitution [25]. Furthermore, the sharpness of the temperature dependence of the magnetization at $T_C$ indicates the high quality of our flux grown $MnSi_{0.992}Ga_{0.008}$ crystals.

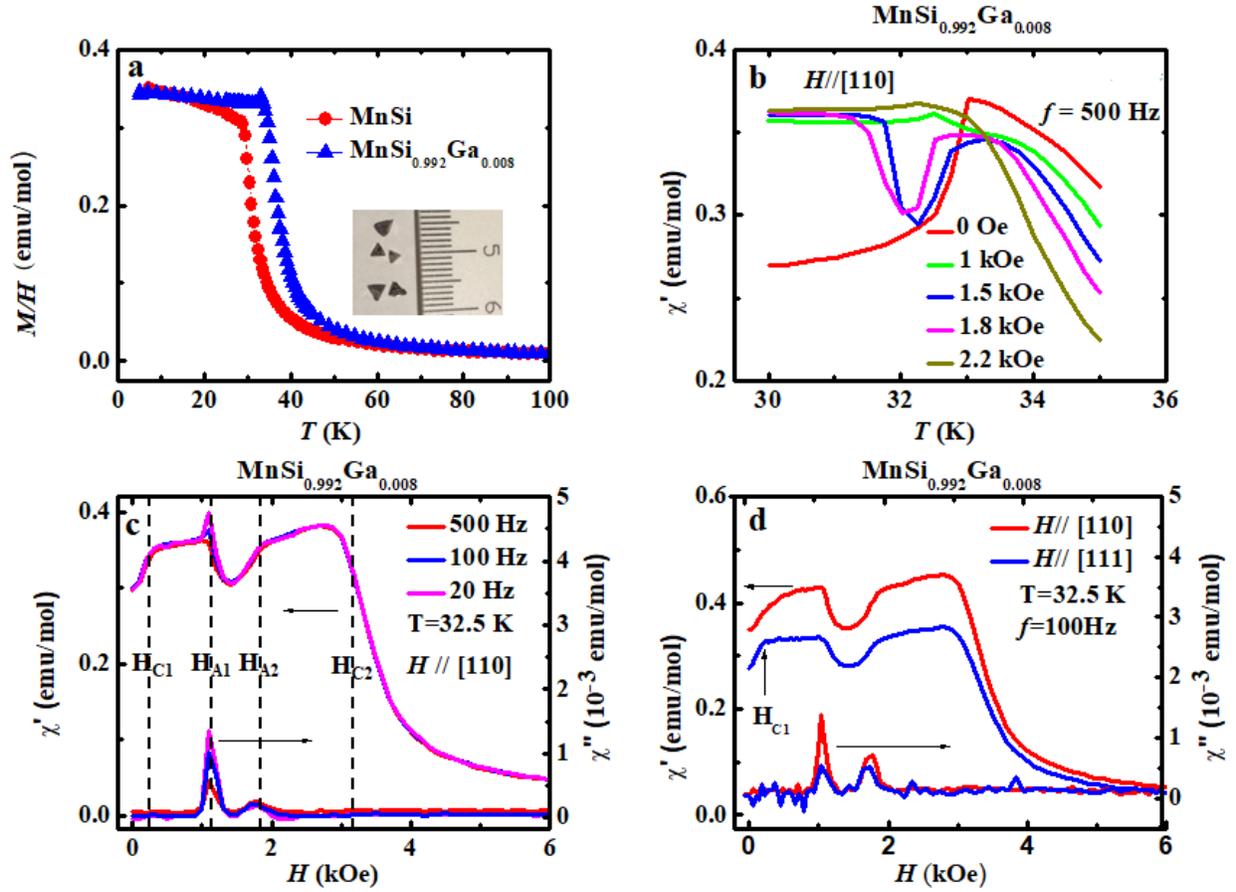

Fig.1 ac and dc magnetization of MnSi$_{0.992}$Ga$_{0.008}$ (a) Temperature, $T$, dependence of the magnetization, $M$ divided by the field, $H$, $M/H$, for MnSi and MnSi$_{0.992}$Ga$_{0.008}$ measured at $H$=1 kOe. Inset: pyramidal shaped crystals grown using Ga flux. (b) Real part of the ac susceptibility, χ', as function of magnetic field, $H$, and $T$. (c) Frequency, $f$, dependence of χ' and the imaginary part of the ac susceptibility, χ", at 32.5 K. (d) Orientation and field dependence of χ' and χ" at 32.5 K. A driving ac field of 2 Oe amplitude was applied during the ac susceptibility measurements.

Fig. 1b presents the real part of the ac susceptibility, χ', for different values of applied dc magnetic field applied along the [110] direction of the crystal. For a range of magnetic fields above 1 kOe, there is a dip in the ac susceptibility just below $T_C$. The span of reduced susceptibility represents the field dependent extent of the SKL. Fig. 1c presents the frequency and magnetic field dependence of χ' and the imaginary part of the ac susceptibility, χ". Such variations of χ' and χ" are typical of all skyrmion hosting $B20$ compounds [2, 12, 26] and the changes evident indicate transitions to the different magnetic states that make up the phase diagram of MnSi. These states include the helical state at zero field with a continuous change to a mulitdomain conical state for $H<H_{C1}$, a single domain conical state for $H_{C1}<H<H_{A1}$, the SKL or A-phase for $H_{A1}<H<H_{A2}$, and a field polarized phase for $H>H_{C2}$ [29]. There is some frequency dependence apparent at the boundary between the conical and SKL phase ($H_{A1}$) with the magnitude of $\chi''$ decreasing from low to high frequency, while there is very little change with frequency to either $\chi'$ or $\chi''$ at $H_{A2}$. This behavior is typical of the response of large magnetic objects having a characteristic time scale in the range of the frequencies employed for the measurement (100's of hertz) [2]. Fig. 1d presents the orientation dependence of $\chi'$ and $\chi''$. These are qualitatively similar for the two orientations displayed but with slightly different $H_{C1}$ values apparent.

## II. Temperature and field history dependence of orientation of the SKL.

We present the results from the SANS measurements in Fig. 2. In these measurements, the high symmetry crystallographic directions [111], [-1-11], [001] and [110] lie in the detector plane as they are perpendicular to both the incident beam ($k_i$ // [-110]) and $H$ ($k_i$ // $H$) (see Fig. 2a). Here, the crystal orientation was held stationary and only $T$ and $H$ were varied between measurements. It is clear from the scattering pattern shown in Fig. 2a, that the helix orientation (along [111] and equivalent [-1-11]) and the magnitude of wave vector of the helix ($Q \sim 0.037$ Å$^{-1}$) [30] remain unchanged from that found in nominally pure MnSi. However, as the remaining scattering patterns reveal, the orientational order of the SKL is very sensitive to temperature and field history and does not remain well registered to a particular crystallographic direction. This is substantially different from what was observed in MnSi [1]. An important observation is that there is only small variation in the translational order indicated by the FWHM along the radial direction. The skyrmion lattice wavevector, $Q$, as function of $H$ and $T$ is shown in Fig. 3 as well as in the supplementary materials [28]. Such a small variation in the radial peak position and FWHM is typical for MnSi near the phase boundary (in $H$ and $T$) between the conical and SKL phases.

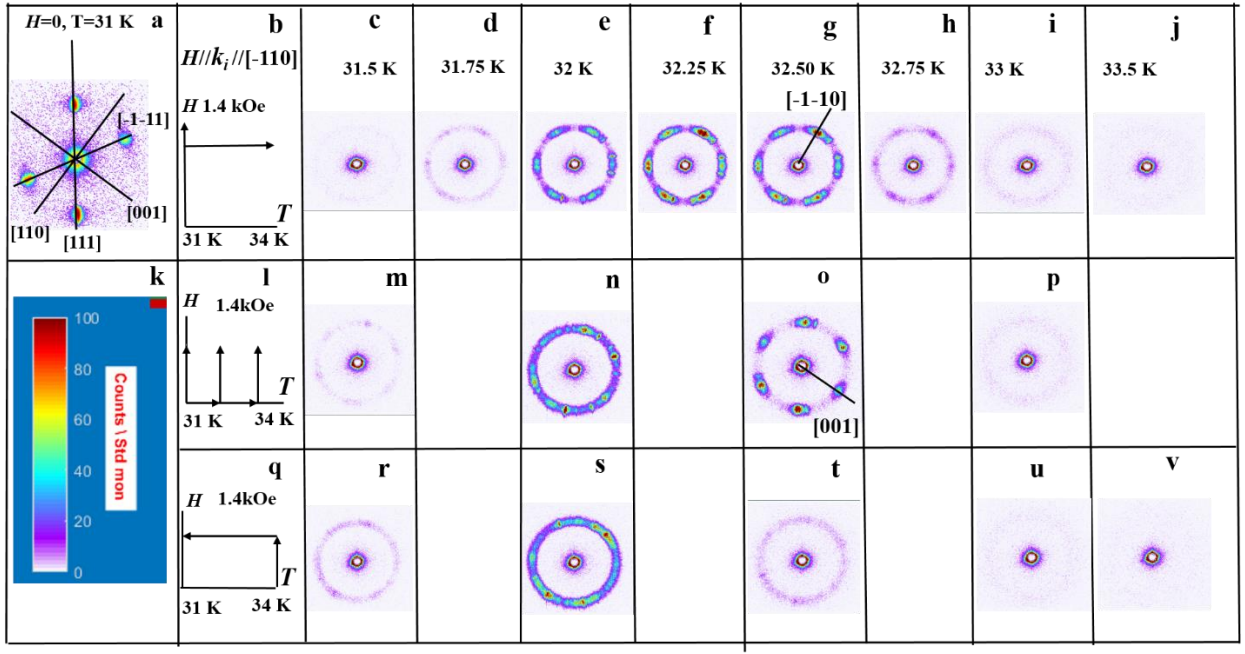

Fig. 2 Small angle neutron scattering (SANS) patterns from MnSi$_{0.992}$Ga$_{0.008}$. (a) Scattering in zero magnetic field, $H$=0, such that the sample is in the helical state. Lines indicate several high symmetry crystallographic directions lying in the detector plane. (b) Summary of $H$ and temperature, $T$, sequences for frames (c) – (j) which present the resulting scattering patterns. (k) Intensity scale. (l) Summary of $H$ and $T$ sequences for frames (m) – (p) which present the resulting scattering patterns. (q) Summary of $H$ and $T$ sequences for frames (r) – (v) which present the resulting scattering patterns. In (b) a constant field of 1.4 kOe was applied and $T$ was increased from 31 K to 34 K. In (l) $H$ = 1.4 kOe was applied at each $T$ and returned to $H$=0 after each measurement. $T$ was then increased by 0.5 K and $H$ ramped back to 1.4 kOe. In (q) a field of 1.4 kOe was applied and the sample cooled from 34 K to 31 K.

We observe that the SKL order can be disrupted, with the scattering forming a ring like feature (see e.g. Fig, 2 n and s), rather than a well-ordered hexagonal pattern, depending upon the $T$ and $H$ history. In addition, for those $T$ and $H$ sequences that yield a well ordered hexagonal pattern indicative of the SKL, the lattice was found to align along either the [110] or [001] so that it appears that neither of these is substantially preferred. We conclude that the tendency for the SKL to orient along a specific crystallographic direction is reduced in MnSi$_{0.992}$Ga$_{0.008}$ compared to MnSi, but that the translational order is retained. The ring-like feature we observe for some sequences indicates that the orientational order becomes either multidegerate, a large number of SKL domains that are randomly oriented are nucleated, or that labyrinth magnetic domains are created [24]. The level of variation that we observe in the $T$ and $H$ history dependent scattering at 32.5+/-0.05 K (Fig. 2 g, o, and t) is surprising.

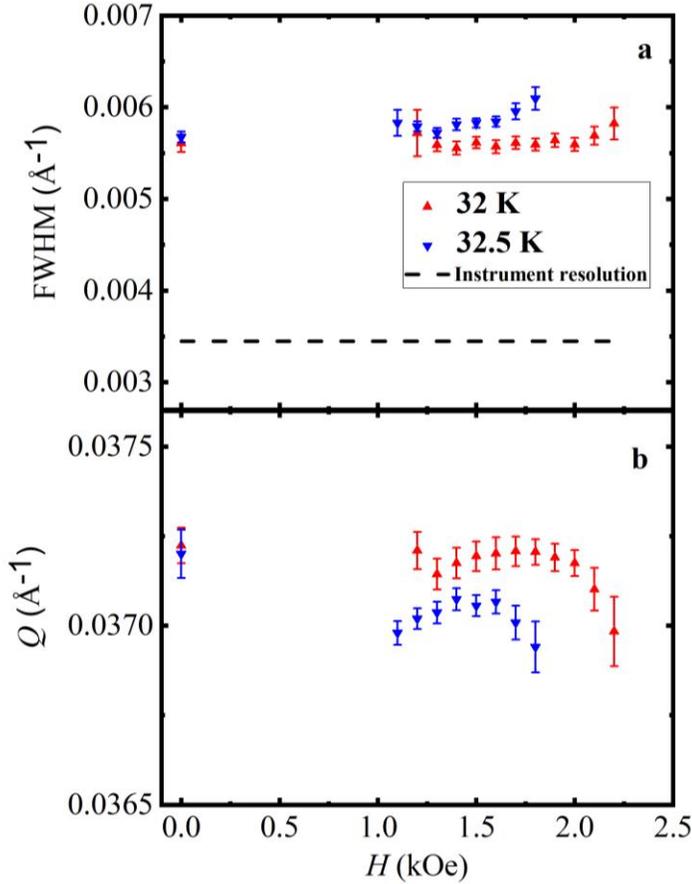

Fig.3 Results of SANS measurement: (a) Variation of full width at half maximum (FWHM) of the radial scans with magnetic field, $H$. Dashed line is the instrumental resolution (~0.0035 Å$^{-1}$) (b) Variation of wave vector $Q$ with magnetic field, $H$. For $0 < H < 1$ kOe, MnSi$_{0.992}$Ga$_{0.008}$ has a conical magnetic order resulting in a loss of scattering intensity on the detector for our experimental geometry. Figs. (a) and (b) share the same symbols.

## III. Magnetic field orientation dependent skyrmion lattice orientation.

In addition to probing the $T$ and $H$ history dependence of the helical and skyrmion states, we have also performed measurements to probe the field orientation dependence as presented in Fig. 4 and in the sequence of figures in the movie in supplementary materials [28]. For these experiments, we began with the crystal oriented similarly to the previous experiments with the crystalline [-110] direction and $H$ aligned along the incident neutron beam and perpendicular to detector plane. The crystalline [-1-11] direction was aligned to be almost vertical whereas in Fig. 2 the vertical direction was along [111]. With this arrangement the reflections associated with the crystalline [110], [001], and [111] directions initially lie in the detector plane. The alignment was within 3° in both horizontal and vertical directions. The crystal was then rotated in the horizontal plane in 5° steps from φ=0 to 90°, as demonstrated in Fig. 4a, such that the [-110] rotates into the detector plane after a 90° rotation. The magnetic field was ramped to zero after each measurement and prior to each crystal rotation and then reenergized to 1.4 kOe.

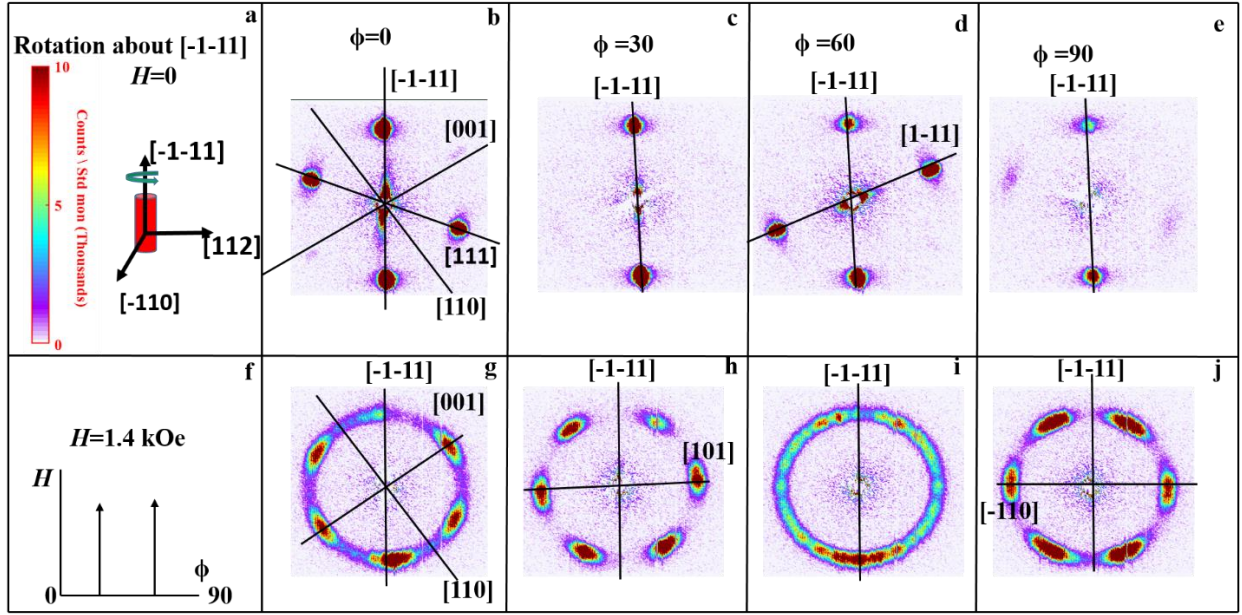

Fig. 4: Field orientation dependence of the skyrmion lattice. SANS data taken as the MnSi$_{0.992}$Ga$_{0.008}$ crystal was rotated about its [-1-11] axis taken at $T$=32.2 K. (a) Intensity scale and schematic of experimental configuration. The red cylinder represents the crystal with three perpendicular axes identified. The crystal was rotated about the vertical axis which was parallel to the crystalline [-1-11] axis. (b) – (e) data taken in zero field, $H$=0. (f) Schematic showing the field and rotation angle, φ, sequence for the measurements presented in frames (g) – (j). $H$ was ramped to zero during the rotation and energized prior to each measurement. (g) – (j) φ dependence of the scattering with $H$=1.4 kOe, within the A-phase of MnSi$_{0.992}$Ga$_{0.008}$. φ, corresponds to the angle measured from the condition that the crystal [-110] direction was aligned parallel to the beam direction and to $H$. In all cases $H$ was parallel to the incident beam and the [-1-11] direction of the crystal was vertical. The [001], [110] and [111] directions lie in the detector plane in frames (b) and (g). Data at a larger number of angles is presented in the sequence of images presented in the supplementary materials [28].

The top row of Fig. 4 displays the scattering at $H$=0 corresponding to the helical state where the crystalline [-1-11] direction is vertical and indicated by the scattering associated with the magnetic domain having $Q$ along this direction. The [-1-11] remained in the detector plane throughout the experiment. Peaks associated with scattering along other [111] equivalent wave vectors progressively appear and disappear as they fall close to the detector plane. The bottom row shows the variation of the scattering pattern at $H$=1.4 kOe where the sample is within the A-phase determined from the magnetic susceptibility data shown in Fig. 1 and confirmed to correspond to a SKL state in both Fig. 2 and Fig. 4. Interestingly, at a rotation angle of 0 the SKL aligns along the crystallographic [001] direction even though the [110] direction is also perpendicular to $H$, thus lying in the detector plane and indicated in Fig. 4b. At $\phi$=30$^o$, the [101] reflection rotates into the detector plane and the SKL aligns along this high symmetry direction. Similarly, for a rotation angle of $\phi$=90$^o$, the [-110] direction rotates into the detector plane and the SKL aligns closely to this direction. Therefore, it appears as if the pinning toward particular crystalline axis is not robust since the SKL evolves so that it aligns to either a [100] (as in Fig. 4g) or a [110] (as in Fig. 4b and 4j) equivalent crystal axis as the crystal is rotated with respect to $H$. If, however, there are no [110] or [100] equivalent crystal axes close to being perpendicular to $H$ (within ~10 degrees), a ring-like scattering pattern results (Fig. 4i). The ring-like structure indicates the lack of an orientationally ordered SKL with the retention of translational order. This is very different from the case of nominally pure MnSi where a six-fold pattern was observed in all cases even when $H$ was applied in an arbitrary direction with respect to the crystal lattice [1, 31]. Furthermore, the radial width of the ring-like feature, which is well above instrumental resolution of ~ 0.0035 Å$^{-1}$, remains nearly the same as that of the hexagonal six-fold patterns [supplementary figure S1[28] and Fig. 3a] indicating an unchanged translational order. In contrast, the azimuthal width continuously varies possibly indicating a change in the orientational order of the SKL with a rotation in the field. The small variation in radial width (Fig.3a) excludes the possibility of formation of a skyrmion glass phase [32].

A comparison of the intensity integrated over $Q$ from 0.028 Å$^{-1}$<$Q$<0.043 Å$^{-1}$ as a function of azimuthal angle $\chi_{az}$ is presented in Fig. 5 for both $H$=0 (Fig. 5a) and for $H$=1.4 kOe (Fig. 5b). Rotation of the crystal with $H$=0 probes the scattering from several different helical domains which are all found to be well aligned along a [111] equivalent crystal direction. In contrast, the orientational order of the A-phase varies dramatically with the orientation of $H$ with respect to the crystalline lattice. The data shown with $\phi$=60$^o$ displays a significant scattering intensity at all $\chi_{az}$ demonstrating a lack of an ordered lattice of spin textures over the volume of the crystal indicating a strong field orientation dependence to the magnetic state.

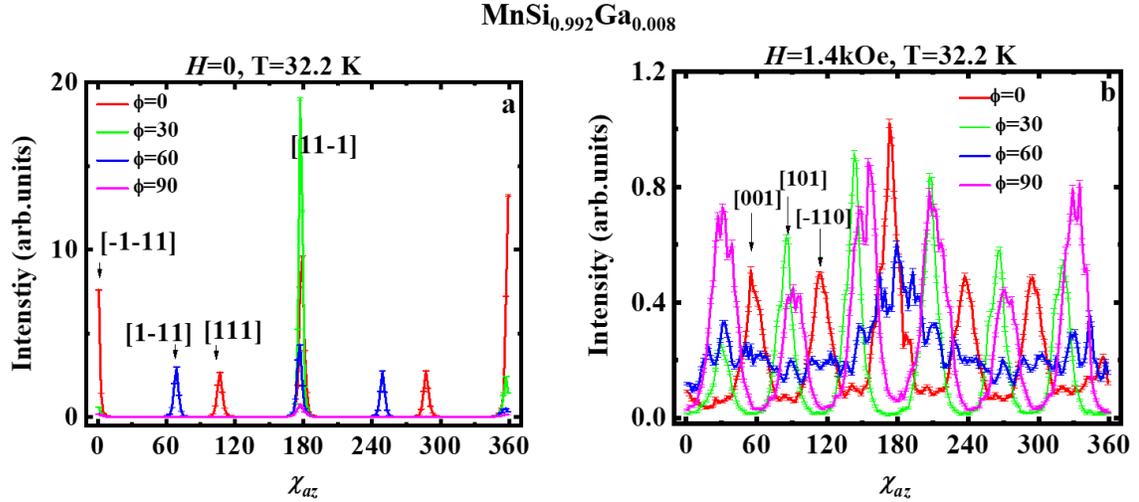

Fig.5 Comparison of $Q$-integrated intensities for (a) zero magnetic field, $H$ (helical state) and (b) $H$=1.4 kOe (A-phase) for different rotation angles, $\phi$, of the crystal with respect to the neutron beam. $\chi_{az}$ is the azimuthal angle in the detector plane. The intensities in both frames are normalized to the same standard monitor count.

## IV. Comparison to Mn$_{1-x}$Ir$_x$Si.

In MnSi it was reported that upon field cycling from $H$=0 to high field and then returning to zero field, there was little change to the scattering patterns or amplitudes [23, 33]. The lack of a history dependent scattering was interpreted in terms of the ease at which the helical domains re-form along each equivalent [111] crystallographic direction after orienting in a magnetic field. This is referred to as the elasticity of reorientation, which reflects the degree that a helical domain remains pinned to a particular [111] axis. We have measured the reorientation process of the helix associated with this elasticity in MnSi$_{0.992}$Ga$_{0.008}$ and Mn$_{0.985}$Ir$_{0.015}$Si single crystals and compared it to that reported for nominally pure MnSi to probe the effect of heavy element substitution on the elasticity of reorientation. This comparison is shown in Fig. 6 where the intensity pattern for the helical state ($H$=0) in both the zero field cooled and field cooled ($H$=5 kOe // [-110]) conditions is presented for both crystals. It is clear that the peaks corresponding to the helical domain lying in the detector plane disappear completely for Mn$_{0.985}$Ir$_{0.015}$Si whereas those peaks reappear with decreased intensity but unchanged orientation for MnSi$_{0.992}$Ga$_{0.008}$. This demonstrates that the elasticity of reorientation of the helix is substantially reduced for Mn$_{1-x}$Ir$_x$Si as compared to MnSi [18]. In addition, we present the scattering for these two samples in a field such that the A-phase is accessed. Figs. 6c and 6f compare the orientation of the SKL for the same orientation of the crystals and magnetic fields after cooling in zero field. In this particular case, the SKL tends to align along the [110] for both $Mn_{0.985}Ir_{0.015}Si$ and for $MnSi_{0.992}Ga_{0.008}$. However, as described above, the orientation of the SKL can switch easily between [001] or [110] for $MnSi_{0.992}Ga_{0.008}$.

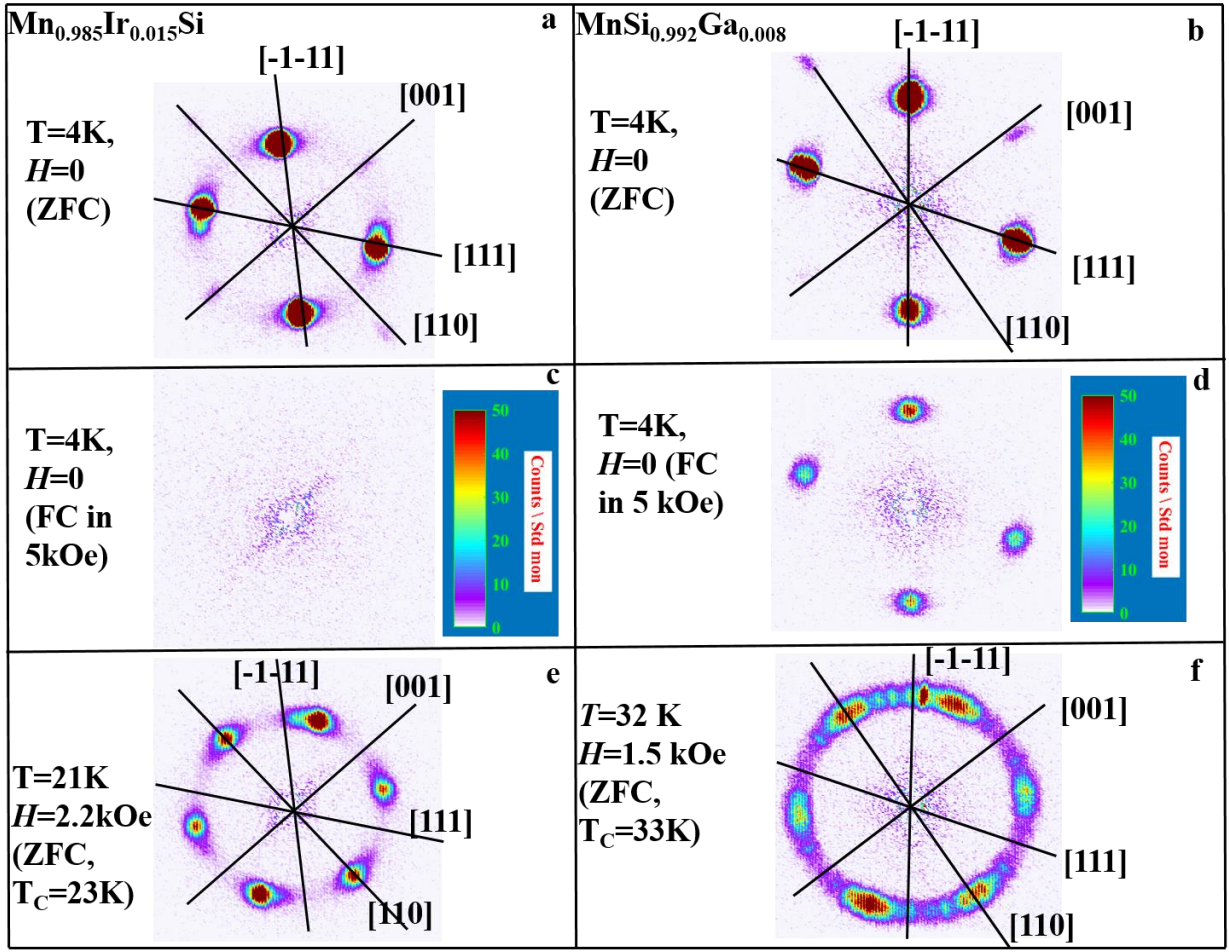

Fig. 6 Comparison of SANS data for $MnSi_{0.992}Ga_{0.008}$ ($T_C$=33.5K) and $Mn_{0.985}Ir_{0.015}Si$ ($T_C$=23 K). Intensity pattern observed in helical state ($T$=4 K, $H$=0) after zero field cooling, ZFC, (a) for $Mn_{0.985}Ir_{0.015}Si$ and (b) for $MnSi_{0.992}Ga_{0.008}$. Intensity pattern observed at $T$=4 K, $H$=0 after field cooling, FC, (c) for $Mn_{0.985}Ir_{0.015}Si$ and (d) for $MnSi_{0.992}Ga_{0.008}$. Intensity pattern in the SKL state (e) for $Mn_{0.985}Ir_{0.015}Si$ and (f) for $MnSi_{0.992}Ga_{0.008}$. Intensity scales are displayed in frames (c) and (d).

## 4. Discussion

The data we present here explores the role of disorder on the formation and orientation of helical and SKL states. Based on these data three important conclusions can be drawn. First, impurities tend to pin helical domains so that field cycling does not return the domains to a state where they are equally distributed along all [111] equivalent directions. This is true for both the Ga and Ir substitution, although the effect appears to be stronger for the heavier, Ir, substitution which at a substitution level of less than twice that of our $MnSi_{0.992}Ga_{0.008}$ crystal results in a null signal above background for the helical domains perpendicular to $H$ in Fig. 6c. This occurs despite the large difference in length scale between the helical wavelength and the atomic scale disorder introduced by the small density of substitutions employed in this work. Second, the SKL in the Ga-substituted system acts in a surprisingly contrasting manner to nominally pure MnSi, as the disorder tends to allow the SKL more freedom to align along different high symmetry directions in the crystal. Third, a disordered state results when the magnetic field is oriented such that there are no high symmetry [100] or [110] axes perpendicular to the field. The disordered A-phase state is somewhat mysterious since the SANS data alone (the ring-like scattering image) are insufficient to define the magnetic state of the system. As pointed out in previous studies [8,17, 23,24,32], the SANS data could be consistent with either a SKL with a wandering orientation as one moves along the magnetic field direction, a SKL with a very large number of randomly oriented domains such that the scattering resembles that of polycrystalline powder, or a labyrinth domain state where the winding number is not constant throughout the crystal.

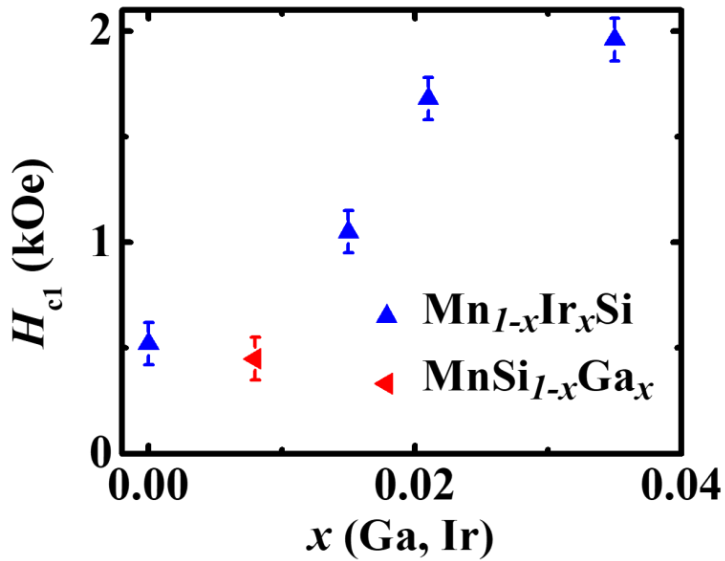

Fig. 7 Variation of $H_{C1}$ for $Mn_{1-x}Ir_xSi$ (blue) and $MnSi_{1-x}Ga_x$ (Red) at 4 K.

We are left with the question of how disorder leads to these changes in the SKL. One possibility is that the substitutions change the local spin-orbit coupling through a change in local electronic structure so that there is a random field acting on the system [34]. The other possibility is that the substitutions cause global change in magnitude and sign of $\in_1$ and $\in_2$ in Eq. 2 that is related to the magneto-crystalline anisotropy [16, 17]. These parameters are thought to determine the orientation of the SKL with respect to the crystal lattice as described in supplementary section of Ref [16]. Furthermore, the spin-orbit interaction is itself responsible for the magneto-crystalline anisotropy causing the moments to align in a particular crystal direction [35]. It is interesting that the disorder created by Ga substitution allows the SKL to align with either the [110] or [001] directions almost interchangeably.

The elasticity of the magnetic lattice depends on the reorientation process of magnetic domains, which involves a length scale much longer than the helical wavelength. In a typical ferromagnet, domains form and reform due to dipolar fields. The cause of formation and reorientation of domains in helimagnets such as MnSi is more complex and not well understood [36-39]. One possibility is that the presence of topological defects act as domain walls as was recently suggested for FeGe [40]. Comparing the reorientation process in $MnSi_{0.992}Ga_{0.008}$ and $Mn_{0.985}Ir_{0.015}Si$ [Fig 6c-6d], one would expect that the disorder associated with Ir likely creates a larger barrier for the reorientation of these domains. The progressive increase of $H_{C1}$ as function of $x$ in $Mn_{1-x}Ir_xSi$ (Fig. 7) in our previous investigation [25,26] also indicates the increase of the barrier potential. It is possible that the difference in the reorientation process of the magnetic states in $MnSi_{0.992}Ga_{0.008}$ and $Mn_{0.985}Ir_{0.015}Si$ is kinetic in origin, relating to the motion of magnetic domains, resulting in a very sluggish return to a multiple domain case when the field is reduced to zero.

To give a final perspective, although, a ring like feature most probably arising from multiple domain SKLs were observed in systems such as $Cu_2OSeO_3$, and (Fe,Co)Si, such a ring like feature was yet to be observed in MnSi [8,17,18,41-46]. The previous observations of ring-like features were thought to be due to the presence of chemical disorder, as well as possible thermal and magnetic field gradients [8, 17, 18, 41-46]. Our investigation shows that even in MnSi with a small level of substitutional disorder where we observe no measurable changes to the helical order, an orientationally disordered SKL can form depending upon the field history (size and orientation) and the temperature. The most likely cause is the increase in internal disorder and/or decreased magneto-crystalline anisotropy caused by the heavier elements substituted for either Si or Mn. Our study provides insight regarding the temperature and field history dependence of the orientation and stability of a SKL in the presence of substitutional disorder. It is not clear at this point, why and how the domains form in helical magnets, how the disorder and topological defects such as dislocations interact with each other, and how these topological defects evolve with field and temperature. This work also establishes the effectiveness of local variations in the spin-orbit interaction to manipulate the SKL despite the difference in length scales between the SKL and the atomic level disorder.

**Note Added**: It was brought to our attention that there are at least 3 related recent manuscripts that address the effect of disorder on the SKL and helical states of B20 structured materials. This includes a SANS exploration of nominally pure MnSi with a rotation of the sample with respect the beam and magnetic field [47] and two papers exploring the effect of Fe and Co substitution in $Mn_{1-x}Fe_xSi$ and $Mn_{1-x}Co_xSi$ [48,49].

**Acknowledgements:** The authors acknowledge R. Jin, W. Xie, D. A. Browne for helpful discussions. We also acknowledge Huibo Cao for discussions and help with the crystal quality check using HB3A beamline in HFIR ORNL. This material is based upon the work supported by the U.S. Department of Energy under EPSCoR Grant No. DE-SC0012432 with additional support from the Louisiana Board of Regents. The SANS data and alignment of the samples used resources at the High Flux Isotope Reactor, a DOE Office of Science User Facility operated by the Oak Ridge National Laboratory. C. Dhital acknowledges Department of Physics Kennesaw State University for the resources provided during preparation of this manuscript.

[*] cdhital@kennesaw.edu

[**] ditusa@phys.lsu.edu

# Unpinning the skyrmion lattice in MnSi; the effect of substitutional disorder

C. Dhital[1,2*], L. DeBeer-Schmitt[3], D. P. Young[1]  and J. F. DiTusa[1**]

*[1] Department of Physics and Astronomy, Louisiana State University, Baton Rouge, LA 70803*

*[2] Department of Physics, Kennesaw State University, Marietta, GA, 30060*

*[3] Oak Ridge National Laboratory, Oak Ridge, Tennessee 37831, USA*

We present more a detailed account of the SANS experiments on $MnSi_{0.992}Ga_{0.008}$ in Figure S1. This includes azimuthally integrated data to compare the $Q$ dependence of the scattering as a function of temperature within in the A, or magnetic skyrmion lattice phase (Fig. S1a) to that in the helical phase at $H=0$ (Fig. S1c).  We have also incorporated plots of the full-width-half-maximum of the $Q$-dependence (Fig. S1b) and the temperature dependence of the magnetic scattering intensity within the helical phase (Fig. S1d). A more complete presentation of the SANS data upon rotation of the crystal about the [-1-11] direction presented in Fig. 4 of the manuscript is included in the sequence of figures contained in Movie 1.

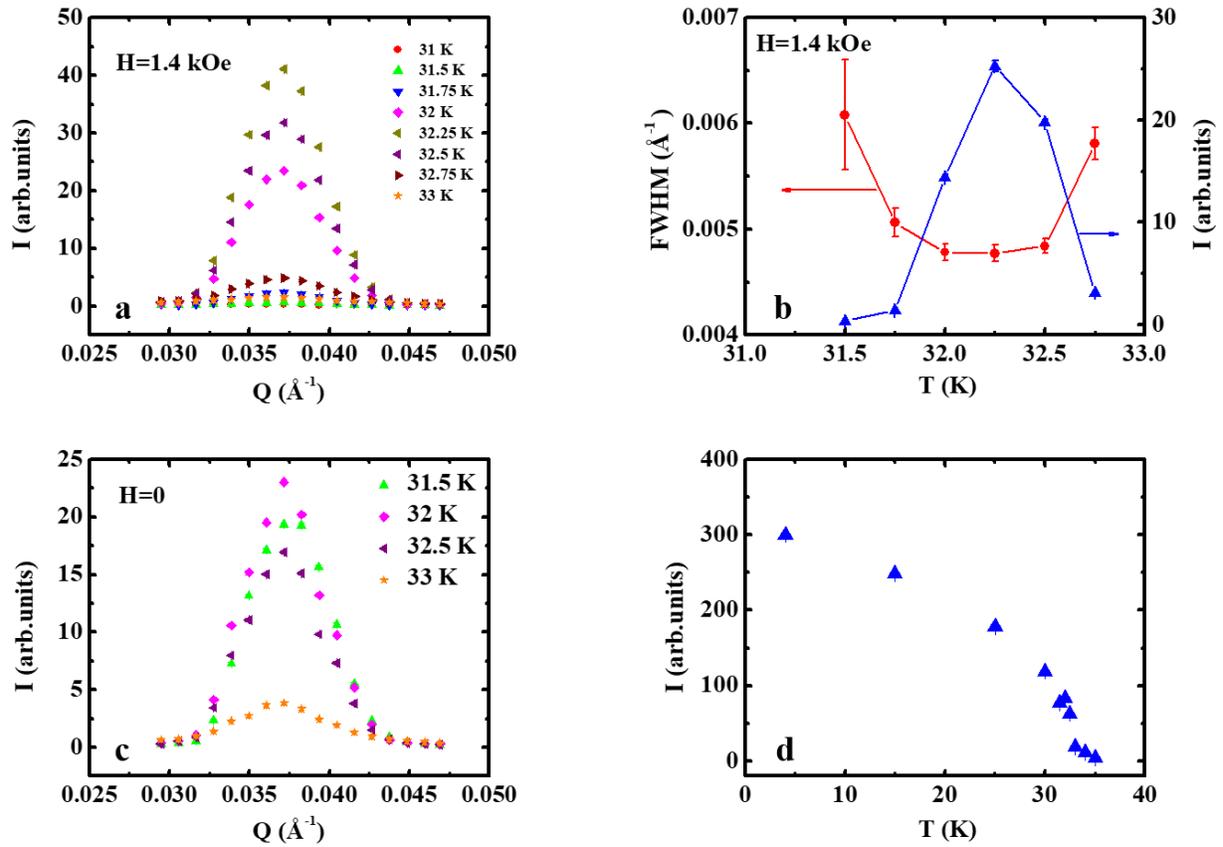

Fig. S1 Small angle neutron scattering on MnSi$_{0.992}$Ga$_{0.008}$ for magnetic field, $H$ // $k_i$ // [-110]. (a) Azimuthally integrated intensity as function of wavevector, $Q$, at temperatures between 31 K and 33 K for $H$=1.4 kOe. (b) Variation of intensity and full-width half-maximum, FWHM, as function of temperature, $T$, for $H$=1.4 kOe. (c) Azimuthally integrated intensity as function of $Q$ at temperatures between 31.5 K and 33 K for $H$=0. (d) Variation of intensity of the helix at $H$=0 as function of temperature.

Movie 1: Scattering pattern as function of rotation angle, $\phi$, for helical phase at zero magnetic field, $H$, and the skyrmion lattice state ($H$=1.4 kOe) at $T$=32.2 K. Here $\phi$ represents the angle between incident beam and [-110] direction. $\phi$ =0 represents $k_i$//$H$//[-110].

MovieOCT_2018.wmv       (wmv format)

MovieOCT_2018.mp4       (mp format)